\documentclass[runningheads]{llncs}
\usepackage[T1]{fontenc}
\usepackage[export]{adjustbox}

\usepackage{graphicx}
\usepackage{caption}
\usepackage{amsmath}
\usepackage{tikz}
\usetikzlibrary{calc}
\usepackage{wrapfig}
\newcommand\T[1]{ \noindent \textbf{#1} }
\usepackage{url}

\usepackage{breakurl}
\usepackage{supertabular}
\usepackage{booktabs}
\usepackage{pifont}%
\usepackage{flushend}
\usepackage{enumitem}%
\usepackage{subcaption}
\usepackage{todonotes}
\usepackage{adjustbox}
\usepackage{float}

\usepackage[subtle,wordspacing=normal, tracking=normal]{savetrees}

\makeatletter
\renewcommand\section{\@startsection{section}{1}{\z@}%
                       {-8\p@ \@plus -4\p@ \@minus -4\p@}%
                       {6\p@ \@plus 4\p@ \@minus 4\p@}%
                       {\normalfont\large\bfseries\boldmath
                        \rightskip=\z@ \@plus 8em\pretolerance=10000 }}
\renewcommand\subsection{\@startsection{subsection}{2}{\z@}%
                       {-8\p@ \@plus -4\p@ \@minus -4\p@}%
                       {6\p@ \@plus 4\p@ \@minus 4\p@}%
                       {\normalfont\normalsize\bfseries\boldmath
                        \rightskip=\z@ \@plus 8em\pretolerance=10000 }}
\renewcommand\subsubsection{\@startsection{subsubsection}{3}{\z@}%
                       {-4\p@ \@plus -4\p@ \@minus -4\p@}%
                       {-1.5em \@plus -0.22em \@minus -0.1em}%
                       {\normalfont\normalsize\bfseries\boldmath}}
\makeatother

\begin{document}

\interlinepenalty=-1
\setlength{\textfloatsep}{5pt}

\setlength{\abovedisplayskip}{3pt}
\setlength{\belowdisplayskip}{3pt}

\title{Dissecting the EIP-2930 Optional Access Lists}

\author{Lioba Heimbach \and Quentin Kniep\and
Yann Vonlanthen\and\\
Roger Wattenhofer \and Patrick Züst}

\institute{ETH Zurich\\Zurich, Switzerland \\
    \email{\{hlioba,qkniep,yvonlanthen,wattenhofer,zuestp\}@ethz.ch}}
\authorrunning{L. Heimbach et al.}
\maketitle              %
\begin{abstract}
Ethereum introduced \textit{Transaction Access Lists (TALs)} in 2020 to optimize gas costs during transaction execution. In this work, we present a comprehensive analysis of TALs in Ethereum, focusing on adoption, quality, and gas savings. 
Analyzing a full month of mainnet data with 31,954,474 transactions, we found that only 1.46\% of transactions included a TAL, even though 42.6\% of transactions would have benefited from it. On average, access lists can save around 0.29\% of gas costs, equivalent to approximately 3,450 ETH (roughly US\$~5~Mio) per year. However, 19.6\% of TALs included by transactions contained imperfections, causing almost 11.8\% of transactions to pay more gas with TAL than without. We find that these inaccuracies are caused by the unknown state at the time of the TAL computation as well as imperfect TAL computations provided by all major Ethereum clients. We thus compare the gas savings when calculating the TAL at the beginning of the block vs. calculating it on the correct state, to find that the unknown state is a major source of TAL inaccuracies. Finally, we implement an ideal TAL computation for the Erigon client to highlight the cost of these flawed implementations. 
\end{abstract}

\section{Introduction}

Ethereum is the second biggest blockchain in terms of market capitalization\footnote{https://coinmarketcap.com (Accessed 20 September 2023)} and the birthplace of \textit{decentralized finance (DeFi)}. DeFi offers many traditional financial services, e.g., exchanges and lending protocols, and has driven activity and demand for block space on the Ethereum blockchain to new levels. 

On Ethereum, users pay for their transaction per unit of \textit{gas}, i.e., a proxy for the computation cost of the transactions. Users further specify how much they are willing to pay per unit of gas. Importantly, higher paying transactions are generally prioritized. As a consequence, the total transaction fees paid by transactions on a blockchain can be seen as a proxy for the demand for block space. Ethereum users currently pay around US\$~3~Mio  in transaction fees per day, overshadowing Bitcoin by a factor of three.\footnote{https://cryptofees.info (Accessed 20 September 2023)} 

Given the immense amount of money spent on Ethereum transactions, it is essential that the computational cost of transactions, i.e., the units of gas charged for the transaction, is estimated correctly. This is also necessary for preventing \textit{denial-of-service (DOS)} attacks on the network. However, this was not always the case and as an indirect consequence, the \textit{Transaction Access List (TAL)} emerged as a solution to contract-breaking issues arising from the needed gas increase. The TAL allows users to specify the addresses and storage keys their transaction will access. They are rewarded for providing the TAL through gas savings but are punished for including too many addresses and storage keys in their TAL. Additionally, we want to highlight that the TAL could be important to guide parallelization down the line~\cite{parallel1,parallel2,parallelflash}. For transactions with perfect TAL, it would be possible to execute transactions with no overlapping entries in parallel, which could increase Ethereum's throughput.

\T{Contribution.} In this work, we present a comprehensive study of the TAL feature from its inception on 15 April 2021 to 31 August 2023. We summarize our contributions as follows: 
\begin{itemize}[topsep=0pt,itemsep=0pt]
    \item We perform a longitudinal study of the usage of TALs to find that while the usage of the TAL is increasing slightly, merely 1.5\% of transactions during our collection period have specified a TAL.
    \item For those transactions that specify a TAL we not only investigate the gas savings but also the causes of increased gas usage in 11.8\% of the cases.  
    \item Our analysis reveals that for around 71\% of all transactions, which could profit from including an ideal TAL, the ideal TAL cannot be computed correctly on the state of the last block. Instead, it requires knowledge on the intra-block state the transaction executes on, which is not known ahead of time --- making a proper TAL computation in these cases nearly impossible. 
    \item Additionally, we find that all major Ethereum clients exhibit flaws in their access list computation. These imperfections can have users paying more when using the TAL computed by their client than without any. 
    \item Finally, we implement a correct TAL computation for the Erigon client, which we made available to Erigon developers and was subsequently implemented~\cite{pullerigon}. Our implementation can also be used as a framework for the other clients. \vspace{2pt}
\end{itemize}

While the TAL was a byproduct of gas increases that led to contract-breaking issues, it also has the potential to improve parallelization by providing additional information for detecting dependencies between transactions prior to execution. We demonstrate though that, due to lack of adoption and correctness, it cannot achieve this in its current state and is likely never will. The TAL feature is available for all users, who may wish to use it to achieve savings on gas fees. Thus, the challenges we present have led to and will continue to lead to unknowing users overpaying for their transactions if not addressed.

\section{The History of Ethereum Transaction Costs}

\label{sec:background}
In Ethereum, the computational complexity of operations is measured in units of gas. Each operation in the \textit{Ethereum Virtual Machine (EVM)} is assigned a fixed amount of gas, representing the computational resources required for its execution.
When users send a transaction to the Ethereum network, they specify the maximal gas amount a transaction is allowed to use, i.e., \textit{gas limit}.
Further, there is a minimum fee charged per transaction which is fixed per block. 

This gas cost mechanism ensures that the financial cost of a transaction is proportional to the computational cost and aims to prevent DOS attacks on Ethereum.
However, historical discrepancies between operation complexity and associated gas costs have led to temporary disruptions in the Ethereum network, requiring adjustments to the gas costs.

\subsection{Changes to Gas Costs in Ethereum}
The Ethereum Yellow Paper~\cite{ethereum_yellow_paper} first assigned gas costs to computational operations in 2014. Subsequently, the first adjustments were made in 2016 as part of \textit{Ethereum Improvement Proposal (EIP)}-150~\cite{eip150}. In this revamp, gas costs for IO-heavy operations were increased to better reflect their actual compute time. This adjustment was preceded by what is known as the ``Shanghai Attack''~\cite{greene2018shanghai}.
Attackers found that certain Ethereum operations, i.e., \texttt{EXTCODESIZE}, \texttt{SELFDESTRUCT}, had inadequate gas costs relative to their required computation time. This led to network congestion and block creation times of more than a minute, effectively bringing the network to a halt.

In 2019, measurements revealed an imbalance between gas cost and computational complexity due to the expanding Ethereum state. The gas cost for affected operations was increased as part of EIP-1884~\cite{eip1884}.

\begin{wraptable}{r}{5.5cm}\vspace{-10pt}
    
        \centering
    \setlength\tabcolsep{0.5em}
    \begin{adjustbox}{width=5.5cm}
    \begin{tabular}{ccc}
        \toprule
        \textbf{Year} & \textbf{Introduced in} & \textbf{Gas cost SLOAD} \\ \midrule
        2014 & Yellow Paper & 20 \\
        2016 & EIP-150 & 200 \\
        2019 & EIP-1884 & 800 \\
        2020 & EIP-2929 & \begin{tabular}{@{}c@{}}2,100 (Cold) \\ 100 (Warm)\end{tabular} \\ \bottomrule
    \end{tabular}
    \end{adjustbox}
    \caption{Gas cost of a storage load (\texttt{SLOAD}) operation over time.}
    \label{tab:sload}\vspace{-10pt}

\end{wraptable}

Then, the most significant overhaul of gas costs came in 2020 after Perez et al.~\cite{perez2019broken} demonstrated that there is generally little correlation between execution cost and utilized resources. An algorithm was presented that generated transactions with a 100x higher execution time than the respective gas cost would suggest. One major problem was that state accesses cost the same independent of whether the respective addresses and storage keys were accessed for the first time in the transaction or had been accessed before (leading to faster cached accesses). EIP-2929~\cite{eip2929} was introduced to address this issue, distinguishing between warm and cold accesses to the Ethereum state. Naturally, the gas cost for cold accesses was set significantly higher than for warm accesses.

We provide an overview of the gas cost development of the \texttt{SLOAD}, i.e., loading a storage cell, instruction in Table~\ref{tab:sload}.
Similar changes were made to the \texttt{*CALL} and \texttt{EXT*} opcode families and the \texttt{BALANCE} and \texttt{SELFDESTRUCT} opcodes.

\subsection{Transaction Access Lists in Ethereum}
Together with the gas increase of EIP-2929, the TAL was proposed and later implemented in EIP-2930~\cite{eip2930}.
This change introduced a new Ethereum transaction variant which optionally includes a parameter called ``accessList'', i.e., a list that contains a set of addresses and mapped storage keys. For each entry in the TAL, an additional gas fee is charged from the sender upfront (2,400 for an address, and 1,900 for each of its storage keys). However, when an address or storage key from the list is accessed for the first time in a transaction, it is already considered warm access instead of cold access. More formally, the respective entries are added to the global sets \texttt{accessed\_addresses} and \texttt{accessed\_storage\_keys} at the start of a transaction execution. The reduced gas costs for subsequent accesses hence compensate for the initial price to include them in the TAL (saving 100 units of gas).
However, the overall gas cost of transactions may increase if non-accessed addresses and storage keys are included in the TAL.

TALs were introduced in 2020 to provide a solution to contract-breaking issues arising from the gas increase in EIP-2929. Such issues emerge when a contract calls another contract with a fixed and hard-coded gas limit. A transaction that worked before the gas cost increase could now revert because the fixed gas limit was chosen too low. By including a TAL, the gas cost of the contract call would decrease, as all accesses are considered warm; mitigating the problem.

Similarly, some contracts using default functions would not work properly anymore after the gas increase. If a contract receives Ether and no payable function is called, a fallback function is executed. This function has a gas stipend of just 2,300 units.\footnote{https://docs.soliditylang.org/en/v0.8.20/security-considerations.html\#sending-and-receiving-ether (Accessed 20 September 2023)} Due to the gas increase, some default functions would now run out of gas while conducting state access operations. This issue could be mitigated by including a TAL, as this would significantly decrease the gas cost of the default function by paying most of the cost for state accesses upfront.

Additionally, TALs could be used for other purposes in the future. Specifically, if all transactions included a perfect TAL, it would be possible to execute transactions with no overlapping entries in parallel. This could increase the throughput of the Ethereum network significantly.

\section{Data Collection}
We ran an Erigon execution layer node and a Lighthouse consensus layer node to collect Ethereum blockchain data. Our data collection window ranges from 15 April 2021 to 31 August 2023. 

We take an in-depth look at August 2023 to pinpoint the achievable gas savings of all transactions with TALs and execute all transactions within this period at least six times, with (possibly) different access lists (including an empty one, the original one, and client-generated ones). To this end, we also modify an Erigon client to allow for an optimized TAL computation.  

\section{Analysis}

\begin{figure}[t]\vspace{-10pt}
    \includegraphics[scale=1]{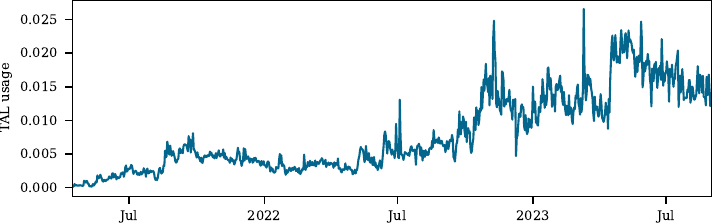}
    \caption{Fraction of transactions per day using the EIP-2930 TALs since their introduction on April 15th 2021 (specifically, at block height 12,244,000).}\label{fig:historical_tal_use}
\end{figure}

We commence by studying the adoption of TALs over time (cf. Figure~\ref{fig:historical_tal_use}). While an initial uptrend in the adoption is visible it appears to have subsided. Further, it can be seen at no point in time was the TAL widely used. 

\begin{table}[b]\vspace{8pt}    
        \centering
        \setlength\tabcolsep{0.5em}
        \begin{adjustbox}{width=\columnwidth}
    \begin{tabular}{@{}lccccc@{}}
        \toprule
        \textbf{Client}        & \textbf{Tx Sender} & \textbf{Tx Recipient} & \textbf{Block Producer} & \textbf{Create} & \textbf{Precompiles} \\ \midrule
        Geth                   &                 &                    & \ding{55}      & \ding{55}       &                      \\
        Nethermind             & \ding{51}       & \ding{51}          & \ding{55}      & \ding{51}       & \ding{55}            \\
        Besu                   & \ding{55}       & \ding{55}          & \ding{55}      & \ding{55}       & \ding{55}            \\
        Erigon                 &                 & \ding{51}          & \ding{55}      & \ding{55}       &                      \\ \midrule
        Ours (modified Erigon) & \ding{51}       & \ding{51}          & \ding{51}      & \ding{51}       & \ding{51}            \\ \bottomrule
    \end{tabular}
    \end{adjustbox}
    \caption{Possible gas optimizations for TALs, i.e., not including addresses that are considered warm (tx sender, tx recipient, block producer, created contracts and precompiles), and whether they are (as of 20 September 2023) implemented by popular Ethereum clients. Empty fields indicate that the case is handled partially. Finally, our modified Erigon client handles all cases. Note that our modifications are now implemented in the Erigon client.}
    \label{tab:optimizations}\vspace{-10pt}         
    
    \end{table}

We now go more into detail on the dataset, which spans all transactions in August 2023 on the Ethereum mainnet. Note that over the month of August 2023, TALs were only included in roughly 1.5\% of transactions.
Importantly, around 20\% of these TALs seen in transactions on the mainnet are suboptimal,
leading to most of these transactions paying more gas fees than they would without providing a TAL.
This can in part be explained by the fact that, as of this writing,
none of the most popular Ethereum clients implement all the optimizations necessary to generate the most gas-efficient TALs (cf. Table~\ref{tab:optimizations}). Common mistakes seen in the clients include the failure to adequately remove the tx sender, tx recipient, and block producer addresses which are considered warm automatically. Further, contracts created in a transaction are warm but often still included in the TAL. Finally, there is a set of warm addresses (i.e., precompiles, Ethereum addresses 0x1 through 0x9) that are considered warm as well but are still often included in TALs. Importantly, including these aforementioned addresses will lead to heightened gas costs. 
Hindering the adoption of the TAL even more, popular ways of interacting with the Ethereum blockchain, such as the Metamask wallet, do not support the feature at all.
Finally, we highlight that it is hard to exactly predict transaction execution before inclusion in a block.
The best we can reasonably expect users to do is execute the transaction on a previous state (ideally at the end of the block immediately prior). Figure~\ref{fig:wrong_tal_use} shows that current TAL users are often unaware of this as many transactions include too many addresses. As indicated by the shading, superfluous addresses are especially costly when they are included in the TALs.

Frustratingly, many redundant addresses do not stem from unpredictable transaction execution, but from mistakenly adding the previously introduced commonly used addresses that do not need to be included as they are considered warm by default. Recall, that adding them into the TAL will still incur the same cost as for any other address but without an upside. Note that this is true, as long as fewer than 24 storage keys are included, in which case the benefits of adding the address and respective storage keys again outweigh the costs.
Table~\ref{tab:suboptimal_addresses} shows that a large portion of mistakes come from such entries, hinting at improper client implementations or uninformed user actions. More than a third of all suboptimal TALs incorrectly include the tx recipient in the TAL, whereas only a few include the block producer and the precompiled addresses. However, this difference might simply stem from these being accessed less frequently overall.

\begin{table}[t]\vspace{-10pt}
\begin{minipage}[b]{.55\linewidth}
    \centering
    \includegraphics[scale=0.94]{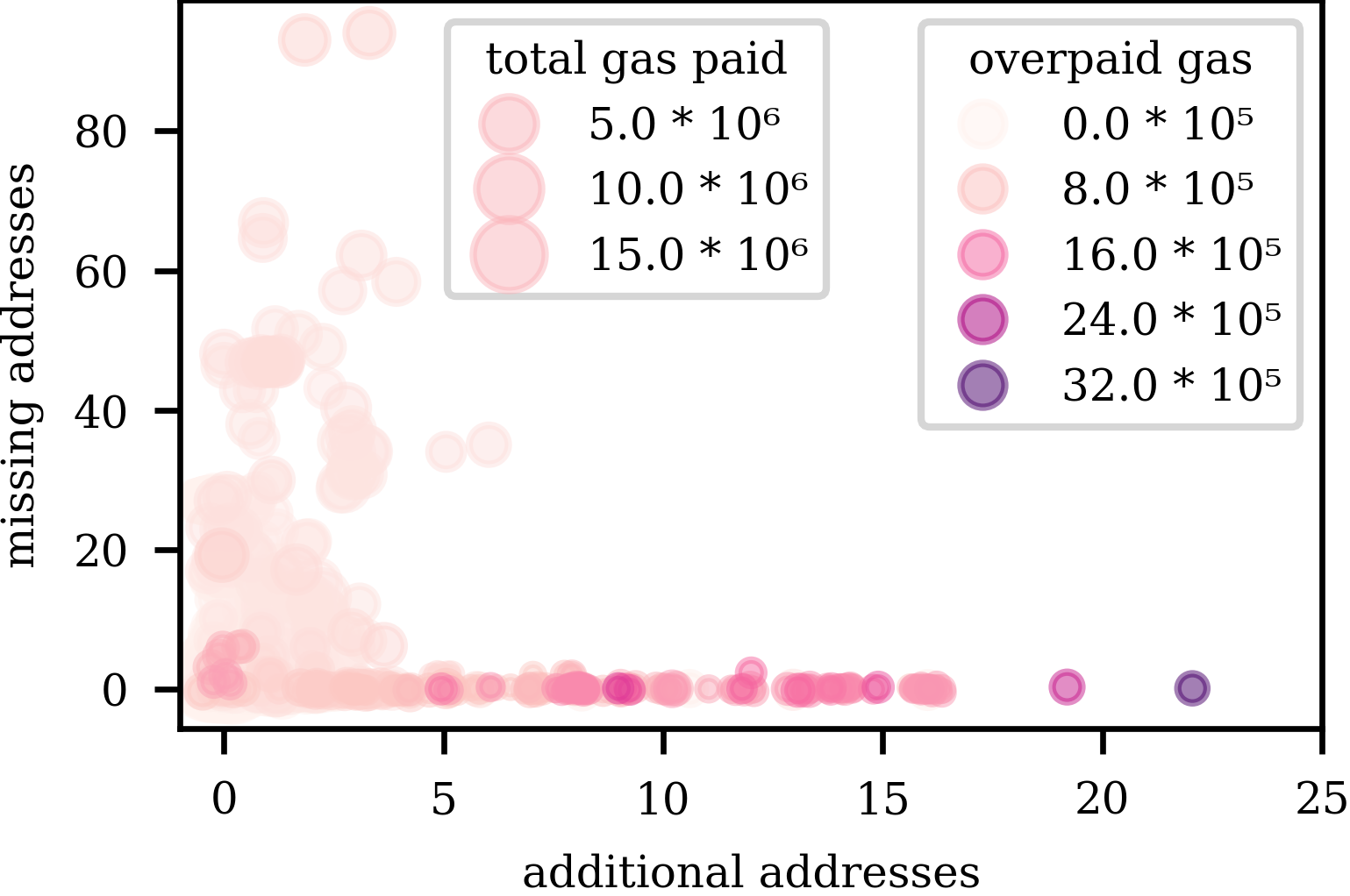}
    \captionof{figure}{Inaccuracies in original TALs.}
    \label{fig:wrong_tal_use}
\end{minipage}\hfill
\begin{minipage}[b]{.42\linewidth}
     \begin{adjustbox}{width=\columnwidth}
    \begin{tabular}{@{}lr@{}}
        \toprule
        \textbf{Address type\hspace{0.5em}} & \textbf{TALs wrongfully including} \\ \midrule
        Block Producer        & 3,710                              \\
        Tx Sender             & 99                                 \\
        Tx Recipient          & 38,942                             \\
        Precompiles           & 2,479                              \\ \bottomrule
    \end{tabular}
    \end{adjustbox}
    \caption{A breakdown of the reasons for suboptimal TALs by 91,255 transactions with inaccuracies. We indicate how many TALs include the given type automatically warm of address, even though they are paying more gas fees doing so.}
    \label{tab:suboptimal_addresses}
\end{minipage}\vspace{-4pt}
\end{table}

\begin{figure}[b]\vspace{4pt}
    \includegraphics[width=\textwidth]{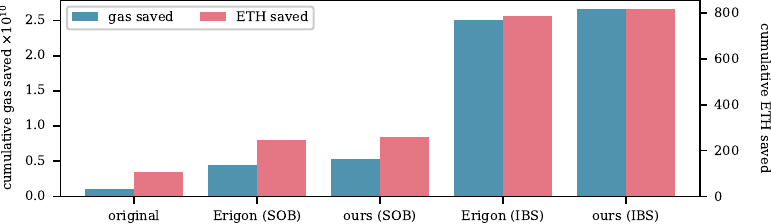}
    \caption{Total transaction fee savings (in gas units and ETH) realized by different access lists,
        all compared to running the transactions without A TAL.
        ``SOB'' refers to generating TALs on the state at the start of the block, whereas ``IBS'' refers to generating TALs on the intra-block state.}\label{fig:tal_profitability}%
\end{figure}

Looking at the profitability of different ways of generating the access lists, we realize that most of the gas savings could only be realized by knowing the exact state the transaction will be executed on (cf. Figure~\ref{fig:tal_profitability}). Unless the transaction creator is also the block producer or pays for bundling, the creator has no certainty of the state on which the transaction will be executed. This has a large impact, as for around 71\% of all transactions that could profit from TALs, generating the TAL at the start of the block is suboptimal.

\begin{figure}[t!]\vspace{-10pt}
    \centering
    \begin{subfigure}[b]{0.48\columnwidth}
        \includegraphics[scale=1,right]{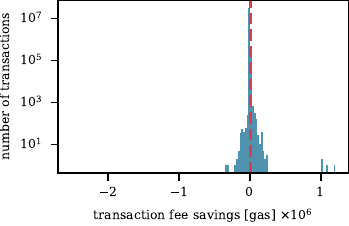}\vspace{0pt}    
    \caption{gas savings of original access list}
    \label{fig:gas_saved_original}
    \end{subfigure}\hfill
    \begin{subfigure}[b]{0.48\columnwidth}
        \includegraphics[scale=1,right]{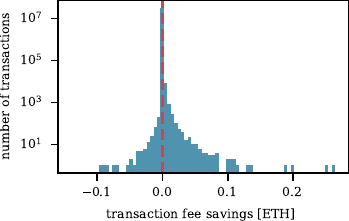}\vspace{2pt}    
        \caption{ETH savings of original access list}
    \label{fig:eth_saved_original}
    \end{subfigure}  \vspace{4pt}

    \begin{subfigure}[b]{0.48\columnwidth}
        \includegraphics[scale=1,right]{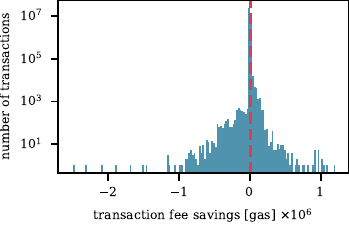}    
     \caption{gas savings of our IBS access list}
    \label{fig:gas_saved_ours_ibs}
    \end{subfigure}\hfill
    \begin{subfigure}[b]{0.48\columnwidth}
        \includegraphics[scale=1,right]{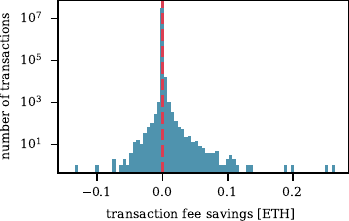}\vspace{2pt}  
    \caption{ETH savings of our IBS access list}
    \label{fig:eth_saved_ours_ibs}
    \end{subfigure}   
    \caption{Distribution of transaction fee savings (in gas units and ETH) realized by the original TALs as well as our optimized TALs generated on the intra-block state (IBS).}
    \label{fig:histo}
\end{figure}

Figure~\ref{fig:histo} visualizes the number of transactions that gain or lose gas from including TALs. Overall, we see a high concentration of transactions where TALs have little impact. In Figure~\ref{fig:eth_saved_ours_ibs} we plot the effect of adding optimal TALs. Note that while upsides are increased, adding a (more precise) TAL can lead to higher gas usage in rare cases, i.e., 0.6\% of all transactions in our analysis. However, this can be explained by transaction behavior changing, e.g., due to parts of the transaction running out of gas later. Nevertheless, in reality, current users seem to be more conservative in their choice of TAL as shown by Figure~\ref{fig:gas_saved_original}.

Finally, we observe that original access lists disproportionately realize gas savings. They reach around a fourth of total potential gas savings (and even over 40\% when measured in ETH), compared to calculating access lists for all transactions at the start of each block, even though these represent just 1.5\% of transactions. This underlines the claim that currently TALs are often used by rather well-versed users on large transactions, that are able to capture a considerable part of benefits that are to be had.

\section{Related Work}
\T{Gas Prices.} Given the immense total cost of Ethereum transactions, an active line of research is devoted to Ethereum gas. One important aspect of Ethereum gas is predicting the next block's price of gas as it is essential for quick block inclusion at fair prices. Multiple works are invested into predicting the price of gas~\cite{werner2020step,mars2021machine,li2021gas,pierro2019influence,pierro2020gas,carl2020ethereum}, as well as understanding the impact on sudden price surges~\cite{faqir2021effect}. 

Given the negative externalities of unpredictable gas prices, EIP-1559 was introduced in 2019. EIP-1559 was shown to be incentive compatible and to stabilize gas prices~\cite{roughgarden2020transaction,leonardos2021dynamical}. Multiple empirical works have further demonstrated the promised improvements in Ethereum transaction fees~\cite{reijsbergen2021transaction,liu2022empirical}. Still, it has been shown that in the presence of farsighted miners/validators~\cite{azouvi2023base,hougaard2023farsighted} rational attacks on the transaction fee mechanism can lead to less predictable fees. 

As opposed to these works, our work does not concentrate on gas prices but on the amount of gas that can be saved by users with the TAL. 

\T{Gas Usage.} Multiple works investigate the usage of gas on Ethereum~\cite{marchesi2020design,albert2020gasol,signer2018gas,li2021gas,zarir2021developing}, e.g., how much gas is used by smart contracts and whether there are possible optimizations to reduce the gas usage. Our work is orthogonal to these, as we focus on the possible gas savings using transaction access lists. 

A separate line of work investigates the Ethereum gas exceptions~\cite{albert2019running,ashraf2020gasfuzzer,liu2020understanding,grech2018madmax}. The TAL, which we investigate, is a byproduct of gas exceptions in various smart contracts, but its impacts were hoped to extend far beyond.

Perez et al.~\cite{perez2019broken} examine the utilized resources of Ethereum transactions to demonstrate that there is generally little correlation between execution cost and utilized resources. As a result of their work, EIP-2929 and TALs were created. To the best of our knowledge, our study is the first to analyze TAL usage and future potential.

\T{Parallelization.} Saraph and Herlihy~\cite{saraph2019empirical} and Chen et al.~\cite{chen2021forerunner} empirically estimate the potential concurrency of speculative execution in Ethereum, while Heimbach et al.~\cite{Heimbach2023defi} quantify the concurrency limits of Ethereum workload assuming accesses are known in advance. In contrast, our work explores to what extent advance knowledge of accesses is realistic.%

\section{Concluding Discussion}

Overall we saw that the potential gas savings are significant when considered in absolute terms.
However, incentives for individual users are small and currently seem insufficient for most users to put in the necessary effort to make proper use of this feature. Even more so, (uninformed) users risk overpaying when including a TAL as a result of flawed TAL computations by all major Ethereum clients and inherent difficulties in proper TAL computations given that the state the transaction executes on is not known ahead of time. A first remedy could arrive from smart contract developers themselves. By providing clear instructions on optimal TAL construction for their developed smart contracts, their users could avoid potential losses, possibly resulting in mutual benefits. 

For now, though, the current lack of adoption also makes TALs virtually unusable for parallelizing transactions. Performing parallel scheduling with (partial) knowledge of dependencies between just 1.5\% of all transactions would have a negligible impact on the overall performance of the network.

Driving the adoption of TALs could be done through mechanism design, for instance by increasing the potential gas savings if a TAL is provided by the user. However, getting the incentives for TALs right could prove to be a difficult task. On the one hand, Ethereum's goal of making gas cost more correlated with computation time would call for incentivizing accesses to warm (cached) data. Yet, to increase the parallelizability of transactions the opposite seems advisable: disincentivizing accesses to frequently used state (at least across transactions).

\newpage

\bibliographystyle{splncs04}
\bibliography{references}

\end{document}